\documentclass[reprint, amsmath,amssymb,showpacs,floatfix,longbibliography, aps, twocolumn, superscriptaddress,nofootinbib]{revtex4-1}
\usepackage{CJKutf8}

\usepackage{amsthm}

\usepackage{multirow}
\usepackage[utf8]{inputenc}
\usepackage{amsmath,amssymb,amsfonts,stmaryrd}
\usepackage{physics}
\usepackage{dsfont}
\usepackage{graphicx}
\usepackage{xcolor}
\usepackage{colortbl}
\usepackage{graphicx}
\usepackage{cancel}
\usepackage{natbib}
\usepackage{color}
\usepackage{tikz}
\usepackage{mathrsfs}
\usepackage{relsize}

\usepackage[all]{xy}
\usepackage{yhmath}
\usepackage{blkarray}
\usepackage{tabularx}
\usepackage{array}
\usepackage{makecell}

\usepackage{diagbox}
\usepackage{algorithm}
\usepackage{algpseudocode}
\usepackage{lipsum}
\newcommand*\colvec[3][]{
    \begin{pmatrix}\ifx\relax#1\relax\else#1\\\fi#2\\#3\end{pmatrix}
}
\algrenewcommand\algorithmicrequire{\textbf{Input:}}
\algrenewcommand\algorithmicensure{\textbf{Output:}}


\usepackage{amsmath}
\usepackage{bm} 
\usepackage{subfigure} 
\usepackage{environ}
\NewEnviron{eqs}{%
\begin{equation}\begin{split}
    \BODY
\end{split}\end{equation}}
\usepackage{url}
\usepackage[margin=0.75in]{geometry}
\usepackage{scalerel}
\usepackage{stackengine,wasysym}

\usepackage{hyperref}
\hypersetup{colorlinks=true,citecolor=blue,linkcolor=blue, urlcolor=blue, breaklinks=true}
\hypersetup{linktocpage}

\allowdisplaybreaks

\usepackage{mathtools}

\newcommand{\ZZ}{{\mathbb Z}}

\graphicspath{{./Figures/}}

\newcommand{\change}{\color{black}}

\usepackage{capt-of}

\begin{document}

\author{Zijian Liang}
\affiliation{International Center for Quantum Materials, School of Physics, Peking University, Beijing 100871, China}
\affiliation{{Hangzhou MatriQ Computing Co., Ltd., Hangzhou 310051, China}}

\author{Yu-An Chen}
\email[E-mail: ]{yuanchen@pku.edu.cn}
\affiliation{International Center for Quantum Materials, School of Physics, Peking University, Beijing 100871, China}

\date{\today}
\title{Generalized $\mathbb{Z}_p$ toric codes as qudit low-density parity-check codes}

\begin{abstract}
We study two-dimensional translation-invariant CSS stabilizer codes over prime-dimensional qudits on the square lattice under twisted boundary conditions, generalizing the Kitaev $\mathbb{Z}_p$ toric code by augmenting each stabilizer with two additional qudits.
Using the Laurent-polynomial formalism, we adapt the Gr\"obner basis to compute the logical dimension $k$ efficiently, without explicitly constructing large parity-check matrices.
We then perform a systematic search over various stabilizer realizations and lattice geometries for $p\in\{3,5,7,11\}$, identifying qudit low-density parity-check codes with the optimal finite-size performance. Representative examples include $[[242,10, {\change \leq} 22]]_3$ and $[[120,6, {\change \leq} 20]]_{11}$, both achieving $k d^{2}/n {\change \leq} 20$.
Across the searched regime, the best observed $k d^{2}$ at fixed $n$ increases with $p$, with an empirical relation $k d^{2} = 0.0541 \, n^{2}\ln p + 3.84 \, n$, compatible with a Bravyi--Poulin--Terhal–type tradeoff when the interaction range grows with system size.

\end{abstract}

\maketitle


\section{Introduction}
Fault-tolerant quantum computation requires active protection against noise, and quantum error-correcting codes are essential to achieve this goal~\cite{Shor1995Scheme, Steane1996Error, Knill1997Theory, gottesman1997stabilizer, kitaev2003fault}.
Topological stabilizer codes, such as toric codes, are promising because they use local parity checks to achieve high thresholds on two-dimensional architectures~\cite{bravyi1998quantum, dennis2002topological, breuckmann2021quantum, google2023suppressing, Google2024surface, liang2025planar}. 
Recent developments on bivariate bicycle (BB) codes show that two-dimensional translation-invariant CSS codes with weight-$6$ checks can outperform the standard toric code on relatively small tori, with performance improved by nearly an order of magnitude~\cite{Bravyi2024HighThreshold, wang2024coprime, voss2024multivariate, berthusen2025toward, eberhardt2024logical, lin2025single}.  
Most of the existing literature, however, focuses on qubit codes, even though several experimental platforms naturally realize higher-level systems such as qutrits or more general qudits~\cite{Goss2022High, Iqbal2025Qutrit, Blok2021Quantum, Morvan2021Morvan, Luo2023Experimental, Cervera2022Experimental, Hu2025Observation, Hu2020Experimental, Liu2023Performing}.  
This motivates the question of how to design and analyze efficient quantum low-density parity-check (LDPC) codes directly in the qudit setting~\cite{Spencer_2026}.

At an abstract level, two-dimensional translation-invariant Pauli stabilizer codes over prime-dimensional qudits that satisfy the topological order (TO) condition
are known to be locally equivalent to stacks of $\mathbb{Z}_p$ toric codes and trivial product states~\cite{bombin_Stabilizer_14, haah_module_13, haah2016algebraic, haah_classification_21, ruba2024homological}.  
The language of topological order and topological quantum field theory (TQFT)~\cite{dijkgraaf1990topological, kitaev2006anyons, ellison2022pauli, liang2023extracting, liang2025operator}, such as anyon types, fusion and braiding, ground-state degeneracy, and Wilson-line operators, can therefore be imported directly into the qudit stabilizer setting.  
From this perspective, a $\mathbb{Z}_p$ qudit stabilizer code can be viewed as a discrete $\mathbb{Z}_p$ topological gauge theory equipped with a built-in notion of topological excitations and their syndromes.

In parallel, an algebraic viewpoint based on Laurent polynomial rings provides a powerful framework for analyzing translation-invariant stabilizer codes~\cite{haah_module_13, haah2016algebraic, liang2023extracting, eberhardt2024logical}.  
By representing Pauli operators as modules over the polynomial ring $\mathcal{R}=\mathbb{Z}_p[x^{\pm1},y^{\pm1}]$, one encodes locality and translation invariance into polynomial data and extracts the associated anyon content via ring-theoretic methods.  
Gr\"obner-basis techniques then enable efficient computation of the ground-state degeneracy and logical operators on twisted tori, without constructing large parity-check matrices.  
In this work, we adapt this machinery to prime-dimensional qudits, focusing on generalized $\mathbb{Z}_p$ toric codes in which the standard $X$-star and $Z$-plaquette stabilizers are augmented by two additional qudits, yielding weight-$6$ checks. 
We then systematically search for low-overhead qudit LDPC codes at system sizes of a few hundred qudits.

\begin{figure*}[t]
    \centering
    \subfigure[Generalized toric codes over $\ZZ_p$.]{\includegraphics[width=0.6\linewidth]{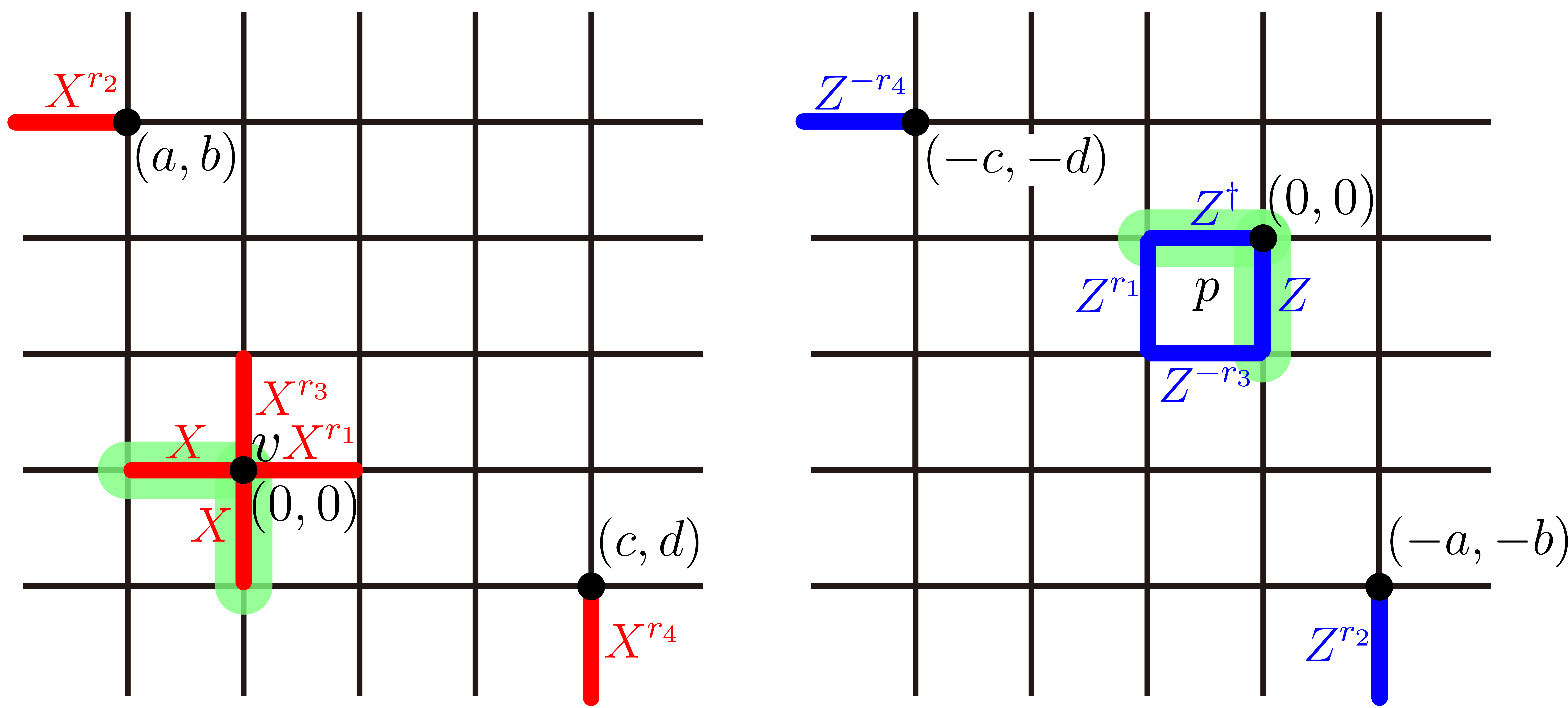}\label{fig: generalized_toric_code}}
    \hspace{2em}
    \subfigure[Twisted torus in three-dimensional space.]{\includegraphics[width=0.35\linewidth]{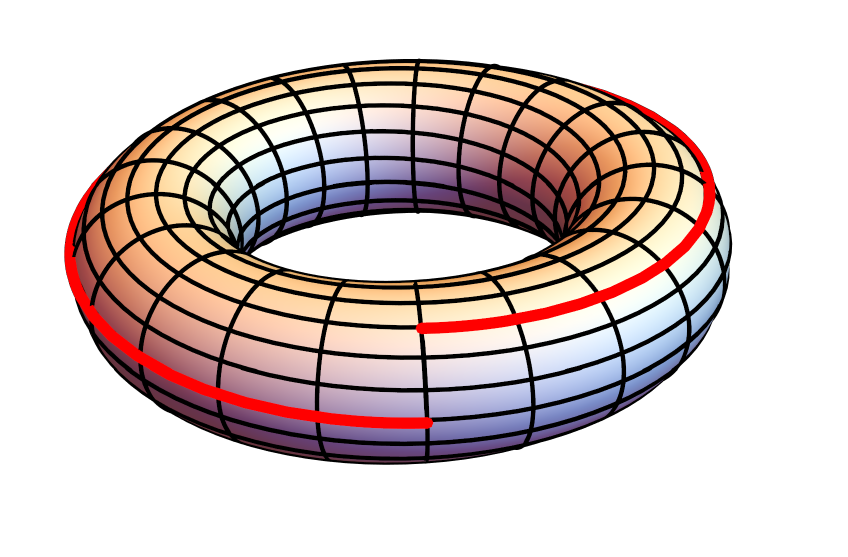}\label{fig: twisted torus}}
\caption{Generalized toric-code stabilizers and twisted boundary conditions.
(a) The $A_v$ and $B_p$ stabilizers of the generalized $\ZZ_p$ toric codes, parameterized by the Laurent polynomials $f(x,y)=1+r_1x+r_2x^ay^b$ and $g(x,y)=1+r_3y+r_4x^cy^d$ in Eq.~\eqref{eq: stabilizer}, with $r_1,r_2,r_3,r_4\in\ZZ_p\setminus\{0\}$ and $a,b,c,d\in\ZZ$. Red and blue edges indicate the supports of the generalized Pauli $X$- and $Z$-type stabilizers, respectively; the labels indicate their Pauli powers. Here $v$ labels a vertex and the subscript of $B_p$ labels a plaquette. The green-shaded edges denote the unit cell at the origin used to generate the Pauli module over the Laurent polynomial ring~\cite{haah_module_13}. For the same pair $(f,g)$, different twisted boundary conditions can realize distinct quantum low-density parity-check codes.
(b) A twisted torus embedded in three-dimensional space, adapted from Ref.~\cite{liang2025Generalized}. The twist is applied along the longitudinal cycle by an angle that is a fraction of $2\pi$, as indicated by the red curve tracing a noncontractible cycle. The torus is specified by the vectors $\vec a_1=(0,\alpha)$ and $\vec a_2=(\beta,\gamma)$: lattice sites obey $\vec v\sim\vec v+\vec a_1\sim\vec v+\vec a_2$.}
    \label{fig: stabilizers and twisted torus}
\end{figure*}

\begin{table*}[t]
\centering
\setlength{\tabcolsep}{3pt}
\begin{minipage}[t]{0.49\textwidth}
\vspace{0pt}
\centering
\textbf{(a)} $\ZZ_{3}$, $n\leq 250$\par\smallskip
{\renewcommand{\arraystretch}{1.25}%
\resizebox{\linewidth}{!}{%
\begin{tabular}{|c|c|c|c|c|c|} \hline
$[[n,k,d]]_3$ & $f(x,y)$ & $g(x,y)$ & $\vec{a}_1$   &$\vec{a}_2$ &$\frac{kd^2}{n}$ \\ \hline
$[[16, 4, 4]]_3$ & $1+x+y$&$1+y-x^{-1}y$ & $(0,4)$&$(2,1)$ &{4} \\ \hline
$[[26, 6, 5]]_3$ & $1+x+x^{-1}y$&$1-y-x^{-1}$ & $(0,13)$&$(1,5)$ &{5.77} \\ \hline
$[[42, 4, 8]]_3$ & $1+x+x^{-2}y^{-1}$&$1+y+x$ & $(0,3)$&$(7,1)$ &{6.10} \\ \hline
$[[48, 4, 9]]_3$ & $1-x+xy^2$&$1-y+x^2$ & $(0,8)$&$(3,3)$ &{6.75} \\ \hline
$[[52, 8, 7]]_3$ & $1+x+x^{-1}y^{-2}$&$1+y+xy^{-1}$ & $(0,13)$&$(2,5)$ &{7.54} \\ \hline
$[[64, 4, 11]]_3$ & $1+x+x^{-1}y^{-2}$&$1+y-x^{-1}y^2$ & $(0,16)$&$(2,7)$ &{7.56} \\ \hline
$[[72, 4, 12]]_3$ & $1-x+y^2$&$1-y+x^2$ & $(0,12)$&$(3,3)$ &{8} \\ \hline
$[[78, 6, 11]]_3$ & $1+x+x^{-1}y^2$&$1-y+x^{-2}y^{-1}$ & $(0,13)$&$(3,4)$ &{9.31} \\ \hline
$[[104, 6, 14]]_3$ & $1-x+xy^{-3}$&$1-y+x^{-2}y^2$ & $(0,26)$&$(2,-9)$ &{11.31} \\ \hline
$[[130, 6, {\change \leq}16]]_3$ & $1+x+x^2y^3$&$1+y-x^{-2}y^2$ & $(0,13)$&$(5,3)$ &{{\change $\leq$}11.82} \\ \hline
$[[144, 6, {\change \leq}17]]_3$ & $1+x+x^2y^2$&$1+y+xy^{-3}$ & $(0,36)$&$(2,15)$ &{{\change $\leq$}12.04} \\ \hline
$[[156, 10, 15]]_3$ & $1+x+x^{-1}y^2$&$1+y+x^{-3}y^2$ & $(0,26)$&$(3,4)$ &{14.42} \\ \hline
$[[160, 12, 14]]_3$ & $1-x-x^{-2}y^2$&$1-y-x^{-3}y^{-2}$ & $(0,40)$&$(2,7)$ &{14.7} \\ \hline
$[[192,8,{\change \leq}19]]_3$ & $1+x+x^2y^{-2}$&$1+y+x^3y$ & $(0,16)$&$(6,7)$ &{{\change $\leq$}15.04} \\ \hline
$[[208,8,{\change \leq}20]]_3$ & $1-x+xy^{-3}$&$1-y+x^{-2}y^2$ & $(0,26)$&$(4,8)$ &{{\change $\leq$}15.38} \\ \hline
$[[224, 6, {\change \leq}24]]_3$& $1+x+x^2y^2$&$1+y+xy^{-3}$ & $(0,56)$&$(2,15)$ &{{\change $\leq$}15.43} \\ \hline
$[[234, 6, {\change \leq}25]]_3$ & $1+x-x^2y^{-6}$&$1+y+x^2y^3$ & $(0,13)$&$(9,2)$ &{{\change $\leq$}16.03} \\ \hline
$[[240, 8, {\change \leq}22]]_3$ & $1+x+xy^{-5}$&$1+y+x^5$ & $(0,60)$&$(2,13)$ &{{\change $\leq$}16.13} \\ \hline
$[[242, 10, {\change \leq}22]]_3$ & $1-x+x^3y^{-3}$&$1-y+x^{-4}$ & $~(0,11)~$&$~(11,4)~$ &{{\change $\leq$}20} \\ \hline
\end{tabular}
}}
\end{minipage}\hfill%
\begin{minipage}[t]{0.49\textwidth}
\vspace{0pt}
\centering
\textbf{(c)} $\ZZ_{7}$, $n\leq 150$\par\smallskip
{\renewcommand{\arraystretch}{1.25}%
\resizebox{\linewidth}{!}{%
\begin{tabular}{|c|c|c|c|c|c|} \hline
$[[n,k,d]]_7$ & $f(x,y)$ & $g(x,y)$ & $\vec{a}_1$   &$\vec{a}_2$ &$\frac{kd^2}{n}$ \\ \hline
$[[12,4,3]]_7$ & $1+2x+5y$   &   $1+2y+2x^{-1}$  & $(0,2)$  & $(3,0)$ &{3.00} \\ \hline
$[[14,4,4]]_7$ & $1+2x+4xy$   &   $1+4y+2xy$  & $(0,7)$  & $(1,3)$ &{4.57} \\ \hline
$[[24,4,6]]_7$ & $1+2x+5y$   &   $1+2y+2x^{-1}$  & $(0,4)$  & $(3,2)$ &{6.00} \\ \hline
$[[28,4,7]]_7$ & $1+5x+x^{-1}y^{-1}$   &   $1+5y+x$  & $(0,7)$  & $(2,3)$ &{7.00} \\ \hline
$[[32,4,8]]_7$ & $1+x+2x^{2}y$   &   $1+2y+3x^{-1}y^{2}$  & $(0,4)$  & $(4,1)$ &{8.00} \\ \hline
$[[48,4,10]]_7$ & $1+x+5x^{2}y$   &   $1+3y+3x^{-1}y^{2}$  & $(0,8)$  & $(3,3)$ &{8.33} \\ \hline
$[[56,4,12]]_7$ & $1+x+5xy$   &   $1+2y+4xy^{-2}$  & $(0,7)$  & $(4,1)$ &{10.29} \\ \hline
$[[64,4,13]]_7$ & $1+2x+2x^{-1}y^{-2}$   &   $1+4y+5xy^{-1}$  & $(0,8)$  & $(4,1)$ &{10.56} \\ \hline
$[[70,4,14]]_7$ & $1+x+5xy^{-2}$   &   $1+4y+2x^{-2}y^{-1}$  & $(0,7)$  & $(5,1)$ &{11.20} \\ \hline
$[[76,6,12]]_7$ & $1+x+x^{-1}y^{2}$   &   $1+2y+4x^{-1}y^{-1}$  & $(0,19)$  & $(2,7)$ &{11.37} \\ \hline
$[[84,4,{\change \leq}16]]_7$ & $1+x+x^{-1}y^{-2}$   &   $1+y+3x^{2}y^{-2}$  & $(0,14)$  & $(3,5)$ &{{\change $\leq$}12.19} \\ \hline
$[[96,6,15]]_7$ & $1+x+y^{2}$   &   $1+4y+5x^{-2}$  & $(0,16)$  & $(3,5)$ &{14.06} \\ \hline
$[[98,10,12]]_7$ & $1+2x+4x^{-1}y^{2}$   &   $1+4y+2x^{-2}y^{-1}$  & $(0,7)$  & $(7,0)$ &{14.69} \\ \hline
$[[126,6,{\change \leq}18]]_7$ & $1+x+x^{2}y^{2}$   &   $1+3y+3x^{3}$  & $(0,21)$  & $(3,15)$ &{{\change $\leq$}15.43} \\ \hline
$[[128,6,{\change \leq}19]]_7$ & $1+x+5x^{3}y^{-2}$   &   $1+2y+4xy^{-4}$  & $(0,16)$  & $(4,7)$ &{{\change $\leq$}16.92} \\ \hline
$[[144,6,{\change \leq}21]]_7$ & $1+x+3x^{2}y^{-2}$   &   $1+y+5x^{-1}y^{-2}$  & $(0,12)$  & $(6,3)$ &{{\change $\leq$}18.38} \\ \hline
\end{tabular}
}}
\end{minipage}

\par\vspace{1em}
\noindent
\begin{minipage}[t]{0.49\textwidth}
\vspace{0pt}
\centering
\textbf{(b)} $\ZZ_{5}$, $n\leq 150$\par\smallskip
{\renewcommand{\arraystretch}{1.25}%
\resizebox{\linewidth}{!}{%
\begin{tabular}{|c|c|c|c|c|c|} \hline
$[[n,k,d]]_5$ & $f(x,y)$ & $g(x,y)$ & $\vec{a}_1$   &$\vec{a}_2$ &$\frac{kd^2}{n}$ \\ \hline
$[[16,4,4]]_5$ & $1+x+3y$   &   $1+3y+x^{-1}y$  & $(0,4)$  & $(2,1)$ &{4.00} \\ \hline
$[[20,4,5]]_5$ & $1+2x+4x^{-1}y$   &   $1+3y+4x$  & $(0,5)$  & $(2,1)$ &{5.00} \\ \hline
$[[24,4,6]]_5$ & $1+2x+2xy$   &   $1+3y+x^{-1}y$  & $(0,4)$  & $(3,2)$ &{6.00} \\ \hline
$[[30,4,7]]_5$ & $1+x+3x^{2}y^{-1}$   &   $1+2y+2x^{2}y$  & $(0,15)$  & $(1,6)$ &{6.53} \\ \hline
$[[40,4,9]]_5$ & $1+x+3x^{2}y^{-1}$   &   $1+3y+x^{2}y$  & $(0,4)$  & $(5,1)$ &{8.10} \\ \hline
$[[48,4,10]]_5$ & $1+x+4x^{-1}y^{-1}$   &   $1+y+2xy^{-1}$  & $(0,6)$  & $(4,3)$ &{8.33} \\ \hline
$[[56,4,11]]_5$ &$1+2x+2x^{-2}$   &   $1+2y+2x^{-1}y^{-1}$  & $(0,4)$  & $(7,1)$ &{8.64} \\ \hline
$[[62,6,10]]_5$ & $1+2x+4y^{2}$   &   $1+4y+3x^{2}y$  & $(0,31)$  & $(1,7)$ &{9.68} \\ \hline
$[[78,10,9]]_5$ & $1+2x+2x^{2}y^{2}$   &   $1+3y+x^{-3}$  & $(0,39)$  & $(1,5)$ &{10.38} \\ \hline
$[[80,6,12]]_5$ & $1+x+3x^{-2}$   &   $1+3y+xy^{-2}$  & $(0,10)$  & $(4,4)$ &{10.80} \\ \hline
$[[90,4,{\change \leq}16]]_5$ &$1+x+x^{-2}y$   &   $1+y+x^{-1}y^{3}$  & $(0,9)$  & $(5,1)$ &{{\change $\leq$}11.38} \\ \hline
$[[96,6,14]]_5$ & $1+2x+2x^{2}y$   &   $1+2y+3x^{-1}y^{2}$  & $(0,16)$  & $(3,6)$ &{12.25} \\ \hline
$[[114,4,{\change \leq}19]]_5$ & $1+x+x^{-1}y^{-2}$   &   $1+y+3x^{3}$  & $(0,19)$  & $(3,7)$ &{{\change $\leq$}12.67} \\ \hline
$[[120,6,{\change \leq}17]]_5$ & $1+x+3y^{3}$   &   $1+2y+2xy^{3}$  & $(0,30)$  & $(2,13)$ &{{\change $\leq$}14.45} \\ \hline
$[[124,6,{\change \leq}18]]_5$ & $1+x+3x^{3}y^{-2}$   &   $1+3y+4x^{3}y$  & $(0,31)$  & $(2,12)$ &{{\change $\leq$}15.68} \\ \hline
$[[144,6,{\change \leq}20]]_5$ & $1+x+3x^{-2}y$   &   $1+y+3x^{2}y^{3}$  & $(0,9)$  & $(8,1)$ &{{\change $\leq$}16.67} \\ \hline
\end{tabular}
}}
\end{minipage}\hfill%
\begin{minipage}[t]{0.49\textwidth}
\vspace{0pt}
\centering
\textbf{(d)} $\ZZ_{11}$, $n\leq 150$\par\smallskip
{\renewcommand{\arraystretch}{1.25}%
\resizebox{\linewidth}{!}{%
\begin{tabular}{|c|c|c|c|c|c|} \hline
$[[n,k,d]]_{11}$ & $f(x,y)$ & $g(x,y)$ & $\vec{a}_1$   &$\vec{a}_2$ &$\frac{kd^2}{n}$ \\ \hline
$[[16,4,4]]_{11}$ & $1+7x+4xy$   &   $1+10y+3x^{-1}y^{2}$  & $(0,4)$  & $(2,1)$ &{4.00} \\ \hline
$[[20,4,5]]_{11}$ & $1+x+2xy$   &   $1+7y+10xy^{-1}$  & $(0,5)$  & $(2,1)$ &{5.00} \\ \hline
$[[22,4,6]]_{11}$ & $1+5x+5x^{-1}$   &   $1+8y+2xy^{-1}$  & $(0,11)$  & $(1,3)$ &{6.55} \\ \hline
$[[32,4,8]]_{11}$ & $1+3x+10y^{3}$   &   $1+10y+3x^{-1}y^{2}$  & $(0,8)$  & $(2,1)$ &{8.00} \\ \hline
$[[40,4,10]]_{11}$ & $1+x+10x^{-1}y^{-1}$   &   $1+7y+10xy^{-1}$  & $(0,10)$  & $(2,4)$ &{10.00} \\ \hline
$[[48,4,11]]_{11}$ & $1+3x+x^{-1}y$   &   $1+6y+x^{-1}y^{-1}$  & $(0,6)$  & $(4,1)$ &{10.08} \\ \hline
$[[60,4,13]]_{11}$ & $1+x+9x^{2}y^{2}$   &   $1+5y+5xy^{-1}$  & $(0,5)$  & $(6,2)$ &{11.27} \\ \hline
$[[66,4,14]]_{11}$ & $1+5x+5y^{2}$   &   $1+8y+2x^{2}$  & $(0,11)$  & $(3,6)$ &{11.88} \\ \hline
$[[72,4,15]]_{11}$ & $1+2x+9x^{-1}y^{-2}$   &   $1+5y+2x^{-2}y^{2}$  & $(0,12)$  & $(3,5)$ &{12.50} \\ \hline
$[[80,6,14]]_{11}$ & $1+x+2x^{-2}y^{-1}$   &   $1+8y+7xy^{-2}$  & $(0,10)$  & $(4,4)$ &{14.70} \\ \hline
$[[96,6,{\change \leq}16]]_{11}$ & $1+2x+8xy^{-3}$   &   $1+3y+7x^{-2}y^{2}$  & $(0,16)$  & $(3,3)$ &{{\change $\leq$}16.00} \\ \hline
$[[100,6,{\change \leq}17]]_{11}$ & $1+x+x^{-2}y^{-1}$   &   $1+5y+10x^{2}y^{-1}$  & $(0,10)$  & $(5,5)$ &{{\change $\leq$}17.34} \\ \hline
$[[120,6,{\change \leq}20]]_{11}$ & $1+x+2x^{-2}y^{3}$   &   $1+y+4x^{-1}y^{-4}$  & $(0,10)$  & $(6,2)$ &{{\change $\leq$}20.00} \\ \hline
$[[144,6,{\change \leq}23]]_{11}$ & $1+2x+8x^{-2}y^{2}$   &   $1+2y+8x^{-2}y^{-3}$  & $(0,8)$  & $(9,1)$ &{{\change $\leq$}22.04} \\ \hline
\end{tabular}
}}
\end{minipage}

\caption{
{\change Qudit LDPC codes over $\ZZ_p$:
(a) $p=3$ with $n\leq250$;
(b) $p=5$, (c) $p=7$, and (d) $p=11$ with $n\leq150$.
The Laurent polynomials $f(x,y)$ and $g(x,y)$ specify the
translation-invariant stabilizers, while the vectors $\vec a_1$
and $\vec a_2$ define the twisted boundary conditions of the torus
(see Fig.~\ref{fig: stabilizers and twisted torus}).
We list $[[n,k,d]]_p$ such that the reported value $k d^{\,2}/n$ for each entry
exceeds those at smaller $n$ within the searched family for the same $p$.
For entries with reported distances at most 15, the code
distances have been determined exactly using the integer
programming approach~\cite{Spencer_2026}.
For every remaining entry, the distance upper bound is obtained
from randomized searches~\cite{Pryadko2022GAP} using a cumulative total of $10^8$
information sets per code.
These distance upper bounds and the corresponding values of
$kd^2/n$ are indicated by the symbol $\leq$.
The distance-estimation and exact-verification procedures are
described in the main text.
Convergence tests for the twelve entries with reported upper
bounds of at least 19 are presented in
Appendix~\ref{app:distance-convergence}.}
}
\label{tab: n_k_d 1}
\end{table*}

\section{Summary of results}
We study two-dimensional translation-invariant CSS codes over $\mathbb{Z}_p$ qudits with $p\in\{3,5,7,11\}$, specified by a pair of Laurent polynomials $f(x,y),\,g(x,y)\in \mathcal{R}$. This polynomial data determines the stabilizers of generalized $\mathbb{Z}_p$ toric codes on twisted tori. 
We systematically enumerate local stabilizer realizations together with twisted boundary conditions, and compute the corresponding code parameters $[[n,k,d]]_p$, where $n$ is the number of physical qudits, $k$ is the number of logical qudits, and $d$ is the code distance. 
For $p=3$ we search system sizes up to $n\le 250$, while for $p\in\{5,7,11\}$ we search up to $n\le 150$. The resulting high-performing qudit LDPC codes and their $[[n,k,d]]_p$ parameters are summarized in Table~\ref{tab: n_k_d 1}.
{\change
The listed instances with reported distances at most 15 have
been verified by exact distance calculations using integer
programming~\cite{Spencer_2026}, with no discrepancies
from the QDistRnd estimates~\cite{Pryadko2022GAP}.
For every listed instance with a reported distance upper bound
greater than 15, we use a cumulative search budget of $10^8$
information sets per code.
Convergence tests for the twelve instances with reported bounds
of at least 19 are presented in
Appendix~\ref{app:distance-convergence}.
The distances not certified by exact calculations remain
explicitly identified as upper bounds.
}

\begin{figure*}[t]
    \centering
    \subfigure[$k d^{2}/n$ vs.\ $n$ for the $\lbrack \lbrack n, k, d \rbrack \rbrack_p$ codes in Tables~\ref{tab: n_k_d 1}--\ref{tab: n_k_d 5}.] {\raisebox{0.27cm}{\includegraphics[width=0.62\linewidth]{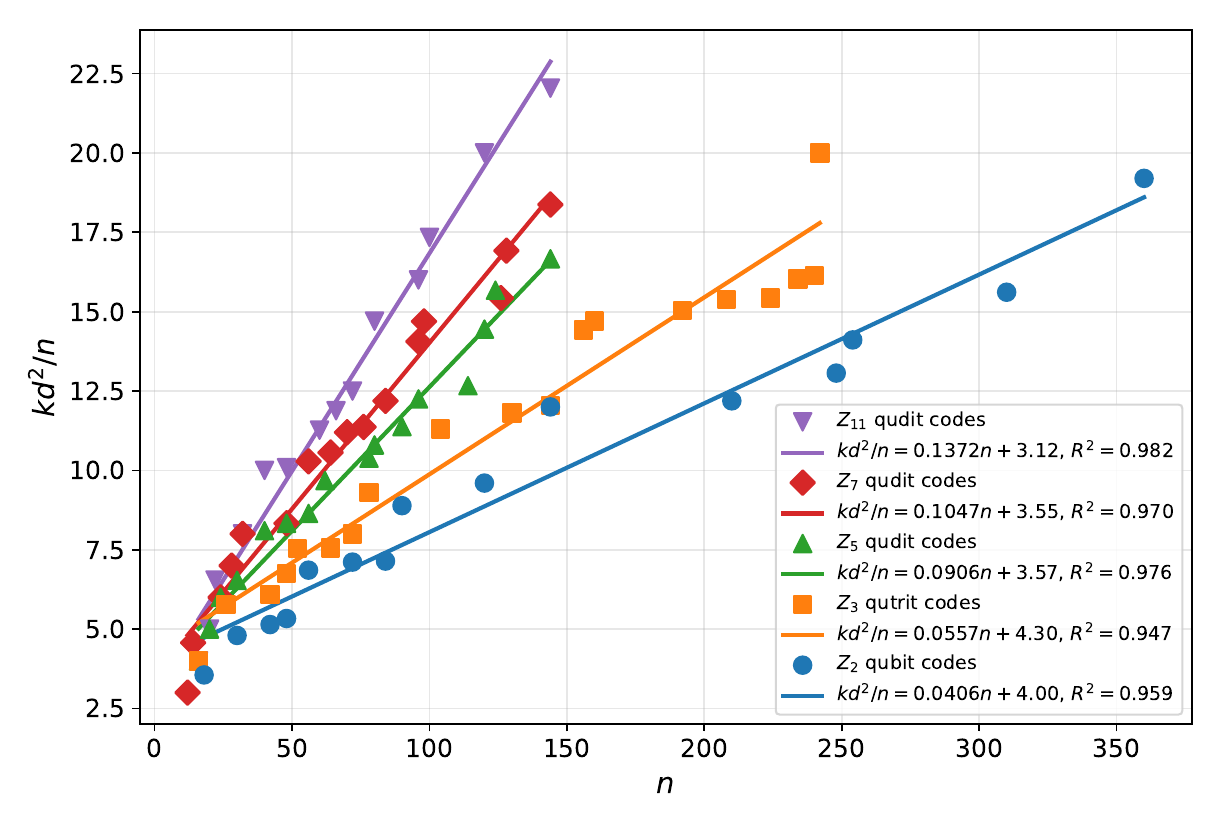}\label{fig: linear fit}}}
    \hspace{-0.75em}
    \subfigure[Fitted slopes from panel (a) vs.\ $\ln(p)$.]{\includegraphics[width=0.38\linewidth]{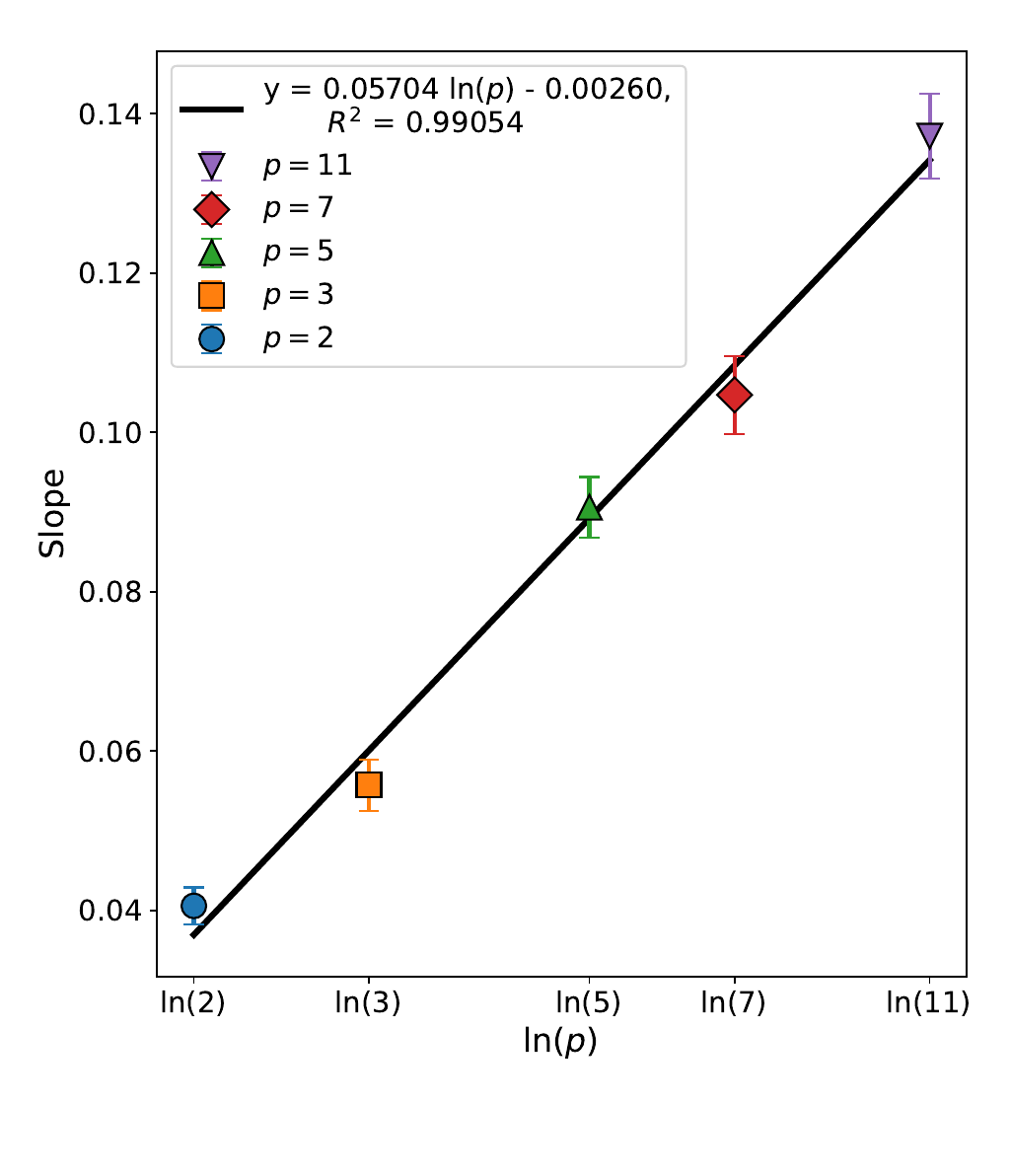}\label{fig: Slope_p}}
\caption{Finite-size performance across qudit dimensions.
(a) The metric $kd^2/n$ for the codes in
Tables~\ref{tab: n_k_d 1}--\ref{tab: n_k_d 5},
plotted against the number of physical qudits $n$.
Blue circles, orange squares, green upward triangles,
red diamonds, and purple downward triangles correspond to
$p=2,3,5,7,11$, respectively; lines of the same colors are
linear fits.
Here $k$ is the number of logical qudits and $d$ is the code
distance, using reported upper bounds where exact distances
are unavailable.
$R^2$ is the coefficient of determination.
(b) Slopes extracted from panel (a), plotted against $\ln p$,
with the same color and marker assignments.
Error bars are the standard errors from the linear fits in
panel (a).
The black line is a linear regression, indicating an
approximately linear dependence on $\ln p$.}
    \label{fig: table to figure}
\end{figure*}

We explicitly present these $[[n,k,d]]_p$ codes via their stabilizers (Fig.~\ref{fig: generalized_toric_code}) and the twisted torus (Fig.~\ref{fig: twisted torus}). For fixed parameters $[[n,k,d]]_p$, there are typically many choices of stabilizer polynomials and lattice vectors; in the tables, we highlight realizations with the most local stabilizers, which are most promising for implementation on qudit hardware. Representative examples include
\begin{eqs}
    &[[52,8,7]]_3,\ [[78,6,11]]_3,\ [[156,10,15]]_3,\ [[242,10,{\change \leq}22]]_3,\\
    &[[30,4,7]]_5,\ [[40,4,9]]_5,\ [[78,10,9]]_5,\ [[120,6,{\change \leq}17]]_5,\\
    &[[28,4,7]]_7,\ [[96,6,15]]_7,\ [[98,10,12]]_7,\ [[144,6,{\change \leq}21]]_7,\\
    &[[40,4,10]]_{11},\ [[100,6,{\change \leq}17]]_{11},\ [[120,6,{\change \leq}20]]_{11}.
    \nonumber
\end{eqs}
In particular, the codes $[[242,10,{\change \leq}22]]_3$ and $[[120,6,{\change \leq}20]]_{11}$ both attain $k d^{2}/n {\change \leq} 20$; their stabilizers and twisted tori are shown in Fig.~\ref{fig: [[242,10,22]] code}.
Moreover, within the searched regime the best observed $k d^{2}$ at fixed $n$ increases with $p$, shown as Fig.~\ref{fig: table to figure}, and the fit in Fig.~\ref{fig: kd2_n2_lnp_fit} yields the empirical relation $k d^{2} = 0.0541\, n^{2}\ln p + 3.84\, n$.

\begin{figure*}[htb]
    \centering
    \subfigure[$\lbrack\lbrack242,10,{\change \leq}22\rbrack\rbrack_3$ code.]{\includegraphics[scale=0.05]{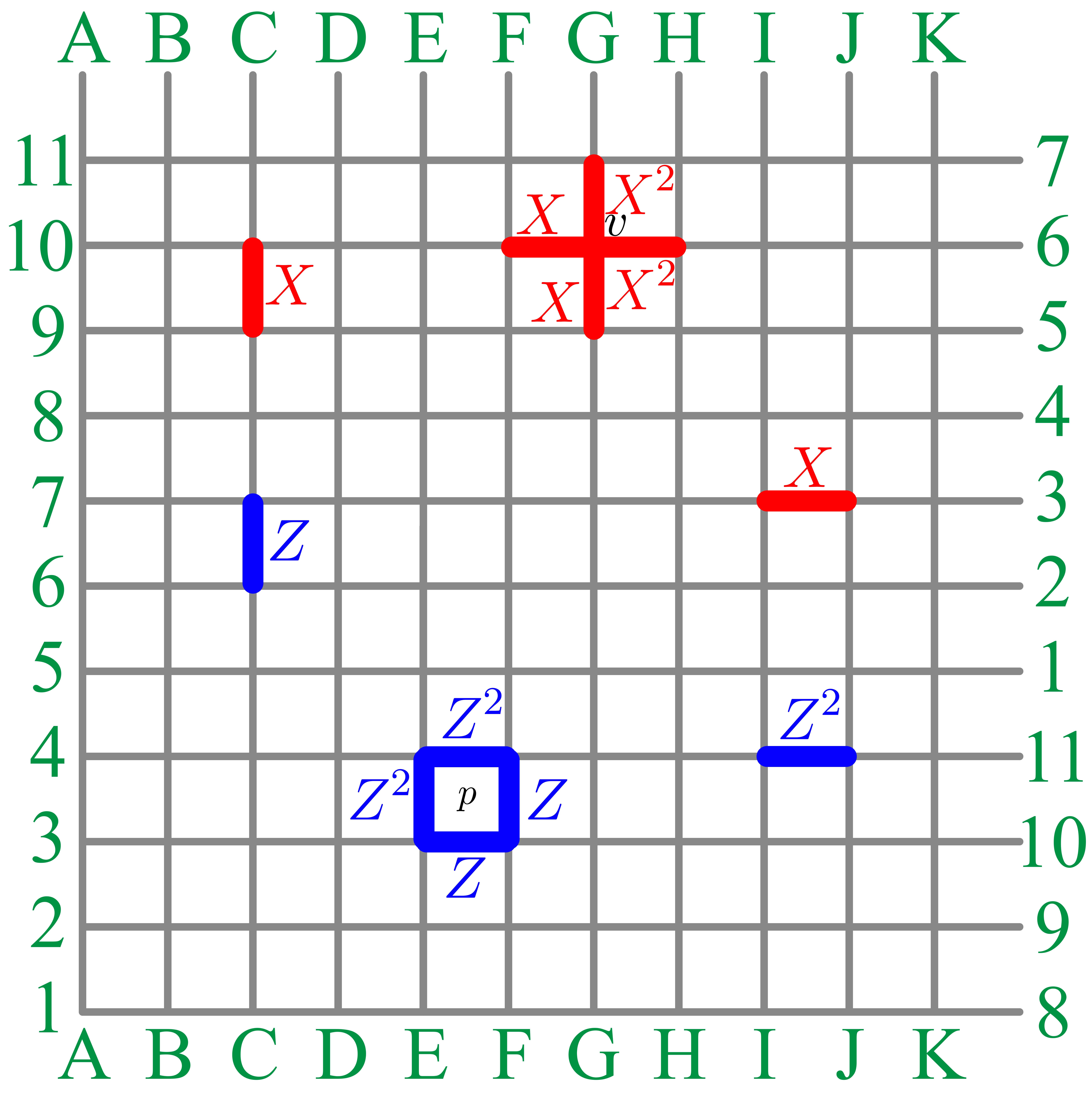}}
    \hspace{2cm}
    \subfigure[$\lbrack\lbrack120,6,{\change \leq}20\rbrack\rbrack_{11}$ code.]{\includegraphics[scale=0.05]{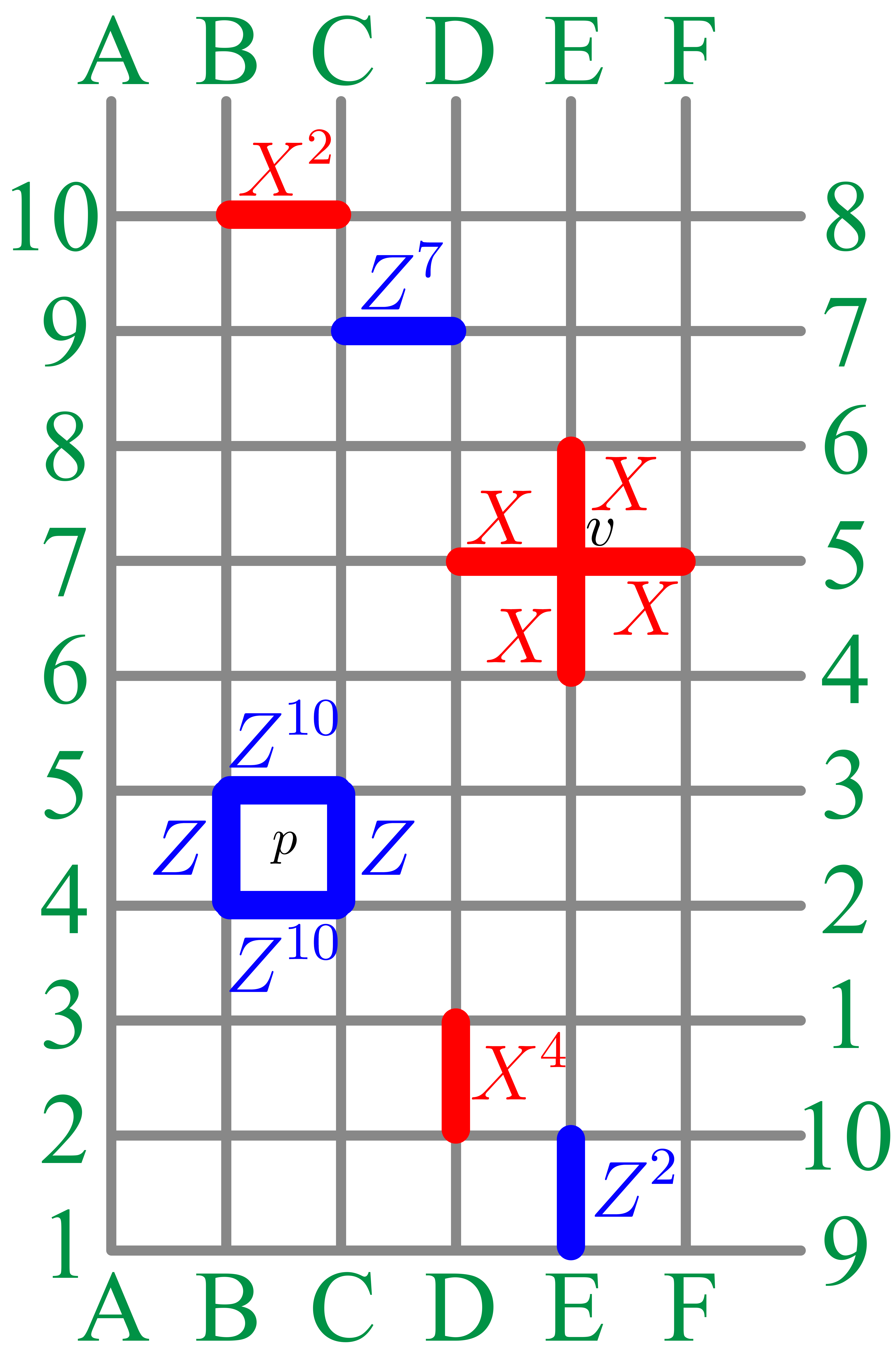}}
\caption{Representative stabilizer realizations on twisted tori.
(a) $X$- and $Z$-stabilizers of the $[[242,10,\leq22]]_3$ code with $f=1-x+x^3y^{-3}$ and $g=1-y+x^{-4}$, defined with twisted boundary vectors $\vec a_1=(0,11)$ and $\vec a_2=(11,4)$.
(b) $X$- and $Z$-stabilizers of the $[[120,6,\leq20]]_{11}$ code with $f=1+x+2x^{-2}y^3$ and $g=1+y+4x^{-1}y^{-4}$, defined with $\vec a_1=(0,10)$ and $\vec a_2=(6,2)$. Red and blue edges show the $X$- and $Z$-stabilizer supports, respectively, and the labels specify the Pauli powers. All other stabilizers are obtained by translations. Vertices with matching green letters on the top and bottom boundaries, or matching green numbers on the left and right boundaries, are identified to form the twisted torus. In $[[n,k,\leq d]]_p$, $n$ and $k$ are the numbers of physical and logical qudits, $p$ is the qudit dimension, and the displayed distance is an upper bound.}
    \label{fig: [[242,10,22]] code}
\end{figure*}

\section{Review of ring-theoretical approach for bivariate bicycle codes}
In this section, we review the algebraic framework for analyzing translation-invariant quantum codes on lattices~\cite{haah2016algebraic}. We follow the notation of Ref.~\cite{liang2025Generalized} and extend its Gr\"obner-basis analysis to $\mathbb{Z}_p$ qudits. 
We begin by recalling the standard $p\times p$ (generalized) Pauli operators for a $\mathbb{Z}_p$ qudit:
\begin{eqs}
    X=\sum_{j \in \mathbb{Z}_p}|j+1\rangle\langle j| ,\quad
    Z=\sum_{j \in \mathbb{Z}_p}  \omega^j |j\rangle\langle j| \text {, }
\end{eqs}
where $\omega$ is defined as $\omega := \exp (\frac{2 \pi i}{p})$. More explicitly,
\begin{eqs}
    X = 
    \begin{bmatrix}
    0 & 0 & \cdots & 0 & 1 \\
    1 & 0 & \cdots & 0 & 0 \\
    0 & 1 & \cdots & 0 & 0 \\
    \vdots & \vdots & \ddots & \vdots & \vdots \\
    0 & 0 & \cdots & 1 & 0
    \end{bmatrix},~
    Z = 
    \begin{bmatrix}
    1 & 0 & 0 & \cdots & 0 \\
    0 & \omega & 0 & \cdots & 0 \\
    0 & 0 & \omega^2 & \cdots & 0 \\
    \vdots & \vdots & \vdots & \ddots & \vdots \\
    0 & 0 & 0 & \cdots & \omega^{p-1}
    \end{bmatrix},
\end{eqs}
and $X$ and $Z$ satisfy the commutation relation
\begin{eqs}
    Z X = \omega X Z.
\end{eqs}
For simplicity, we focus on the square lattice with one $\ZZ_p$ qudit living at each edge. We briefly review the polynomial representation of Pauli operators~\cite{haah_module_13}. A unit cell contains two vertices, $e_1$ and $e_2$, whose Pauli operators are represented by four-dimensional vectors:
\renewcommand{\arraystretch}{1.3}
\begin{equation}
    \mathcal{X}_{e_1} =
    \begin{bmatrix}1 \\ 0 \\ \hline 0 \\ 0\end{bmatrix},~
    \mathcal{Z}_{e_1} =
    \begin{bmatrix}0 \\ 0 \\ \hline 1 \\ 0\end{bmatrix},~
    \mathcal{X}_{e_2} =
    \begin{bmatrix}0 \\ 1 \\ \hline 0 \\ 0\end{bmatrix},~
    \mathcal{Z}_{e_2} =
    \begin{bmatrix}0 \\ 0 \\ \hline 0 \\ 1\end{bmatrix}.
    \label{eq: X1 Z1 X2 Z2 definition}
\end{equation}
A translation by $(n,m)\in\mathbb{Z}^2$ is implemented by multiplying the corresponding vector by the monomial $x^{n}y^{m}$. 
More generally, any Pauli operator (modulo an overall phase) is represented by a vector in $\mathcal{R}^{4}$, where $\mathcal{R}=\mathbb{Z}_p[x^{\pm1},y^{\pm1}]$. 
Vector addition corresponds to operator multiplication, while multiplication by elements of $\mathcal{R}$ encodes lattice translations and $\mathbb{Z}_p$ exponents. 
Thus, the Pauli group modulo phase is naturally identified with an $\mathcal{R}$-module.

A translation-invariant CSS code is specified by a pair of Laurent polynomials $f,g\in \mathcal{R}$:
\renewcommand{\arraystretch}{1.4}
\begin{equation}
    A_v = 
    \begin{bmatrix} f(x,y) \\ g(x,y) \\ \hline 0 \\ 0 \end{bmatrix}, 
    \qquad
    B_p = 
    \begin{bmatrix} 0 \\ 0 \\ \hline -\overline{g(x,y)} \\ \overline{f(x,y)} \end{bmatrix},
    \label{eq: stabilizer}
\end{equation}
where $\overline{(\cdot)}$ denotes the antipode map $x^{n}y^{m}\mapsto x^{-n}y^{-m}$. The stabilizer group is generated by all lattice translates of $A_v$ and $B_p$. 
As a simple example, the Kitaev toric code~\cite{bravyi1998quantum} corresponds to $f(x,y)=1-x$ and $g(x,y)=1-y$. 

The topological order condition is satisfied when $f(x,y)$ and $g(x,y)$ are coprime~\cite{eberhardt2024logical}, which ensures that any local operator commuting with all stabilizers is itself a stabilizer.
Moreover, the maximal number of logical qudits on a torus equals the number of independent anyon generators in the underlying topological order~\cite{Witten1989Jones, Wen1995Topological, watanabe2023ground}. It can be computed from the codimension of the ideal generated by $f$ and $g$:
\begin{equation}
    k_{\mathrm{max}}
    = 2\,\dim\!\left(
    \frac{\mathbb{Z}_p[x^{\pm1},y^{\pm1}]}{\langle f,\,g\rangle}
    \right).
    \label{eq: maximal logical dimension}
\end{equation}
For twisted boundary conditions specified by $\vec{a}_1=(0,\alpha)$ and $\vec{a}_2=(\beta,\gamma)$, the resulting code has logical dimension~\cite{liang2025Generalized}
\begin{equation}
    k
    = 2\,\dim\!\left(
    \frac{\mathbb{Z}_p[x^{\pm1},y^{\pm1}]}{\langle f(x,y),\,g(x,y),\,y^{\alpha}-1,\,x^{\beta}y^{\gamma}-1\rangle}
    \right).
    \label{eq: k formula}
\end{equation}
The derivations of Eqs.~\eqref{eq: maximal logical dimension} and~\eqref{eq: k formula} for $\ZZ_p$ qudits follow directly from the corresponding proofs for the $\ZZ_2$ case established previously in Refs.~\cite{liang2025Generalized, chen2025anyon}.
The right-hand side of Eq.~\eqref{eq: k formula} can be evaluated efficiently using a Gr\"obner basis computed via Buchberger’s algorithm~\cite{Buchberger1965algorithm}.

\section{Search for bivariate bicycle codes over $\ZZ_p$}
We introduce the \emph{generalized toric code} over $\ZZ_p$, a particular subclass of bivariate bicycle (BB) codes, defined by
\begin{eqs}
    f(x,y) &= 1 + r_1 x + r_2 x^{a} y^{b}, \\
    g(x,y) &= 1 + r_3 y + r_4 x^{c} y^{d},
\label{eq: a b c d generalized toric code}
\end{eqs}
with $r_i\in\ZZ_p\setminus\{0\}$ and $a, b,c,d \in \ZZ$, as illustrated in Fig.~\ref{fig: generalized_toric_code}.

We systematically search for high-performing weight-$6$ BB codes of the form~\eqref{eq: a b c d generalized toric code} over $\ZZ_p$ qudits with twisted-periodic boundary conditions. For each even $n$, we enumerate all factorizations $n=2\alpha\beta$ and define a twisted torus by the basis vectors $\vec{a}_1=(0,\alpha)$ and $\vec{a}_2=(\beta,\gamma)$ with $0\le \gamma<\alpha$. We then enumerate all polynomials of the form~\eqref{eq: a b c d generalized toric code} whose exponent pairs $(a,b)$ and $(c,d)$ lie within the fundamental parallelogram spanned by $\vec{a}_1$ and $\vec{a}_2$. For each candidate, we compute the corresponding logical dimension $k$ using Eq.~\eqref{eq: k formula}.
The computational cost of evaluating $k$ in Eq.~\eqref{eq: k formula} is essentially independent of system size: the boundary relations $y^{\alpha}-1$ and $x^{\beta}y^{\gamma}-1$ are reduced modulo the ideal $\langle f,g\rangle$, and only the resulting remainders enter the Gr\"obner-basis dimension calculation.

{\change
For instances with $k>0$, we first obtain upper bounds on the
code distance using QDistRnd~\cite{Pryadko2022GAP}. 
For the instances in Table~\ref{tab: n_k_d 1} whose initial upper
bounds are at most 15, we further determine the exact
distances by the integer programming
approach~\cite{Spencer_2026}.
In every case, the exact distance agrees with the initial
estimate.
For every listed instance with a reported upper bound greater
than 15, we perform randomized searches using a cumulative total
of $10^8$ information sets per code, summed across parallel
search streams.
The smallest nontrivial logical-operator weight found is reported
as a distance upper bound.
We additionally assess convergence for the twelve instances
with reported upper bounds of at least 19.
For each of these codes, all 100 independently seeded runs
attain the reported bound within $10^5$ information sets per run.
These results, detailed in
Appendix~\ref{app:distance-convergence}, support the empirical
reproducibility and stability of the reported bounds
but do not certify exact distances.
}

The results for qudit LDPC codes over $\mathbb{Z}_p$ with $p=3,5,7,11$ are summarized in Table~\ref{tab: n_k_d 1}, covering $p=3$ with $n\le 250$ and $p =5,7,11$ with $n\le 150$.
{\change
For each fixed $p$, our search exhaustively enumerates the
allowed polynomial and twisted-boundary parameters for every
even $n$ within the stated search range and the ansatz in
Eq.~\eqref{eq: a b c d generalized toric code}.
Table~\ref{tab: n_k_d 1} presents a selected subset of these
results. At each $n$, we first identify the largest reported
value of $kd^2/n$, and retain a representative only if this
value strictly exceeds those obtained at all smaller $n$.
When several codes at the same $n$ attain this value, we
choose the realization with the most local stabilizers.
The listed values therefore increase by construction.
Codes that do not improve this running maximum are omitted;
their values need not increase monotonically with $n$,
as illustrated by the qubit $[[n,k,d]]_2$ results in
Ref.~\cite{liang2025Generalized}.
The plots and fits below use this selected set.
}

By comparing the $\ZZ_2$ results~\cite{liang2025Generalized} summarized in Appendix~\ref{app: Z2 BB codes} with our data for $\ZZ_3$, $\ZZ_5$, $\ZZ_7$, and $\ZZ_{11}$, we obtain the plot in Fig.~\ref{fig: linear fit}. The vertical axis shows the figure of merit $k d^{2}/n$, while the horizontal axis shows the minimum number of physical qudits $n$ required to achieve a given value of $k d^{2}/n$. 

For each fixed $p$, the dependence of $k d^{2}/n$ on $n$ is approximately linear over the fitted range, with coefficient of determination $R^{2}\approx 0.95$--$0.98$.
Moreover, at fixed $n$ we observe that the best achievable $k d^{2}/n$ increases with the qudit dimension $p$. For example, at comparable performance, the $[[120,6,{\change \leq}20]]_{11}$ code ($k d^{2}/n{\change \leq} 20$) uses $n=120$ qudits, whereas the best $\ZZ_2$ instance in our comparison, $[[360,12,{\change \leq}24]]_2$ ($k d^{2}/n{\change \leq}19.2$), requires $n=360$ qubits. Finally, the linear fits in Fig.~\ref{fig: linear fit} suggest that the fitted slope of $k d^{2}/n$ versus $n$ grows approximately as $\ln p$, as summarized in Fig.~\ref{fig: Slope_p}.

\begin{figure}
    \centering
    \includegraphics[width=1\linewidth]{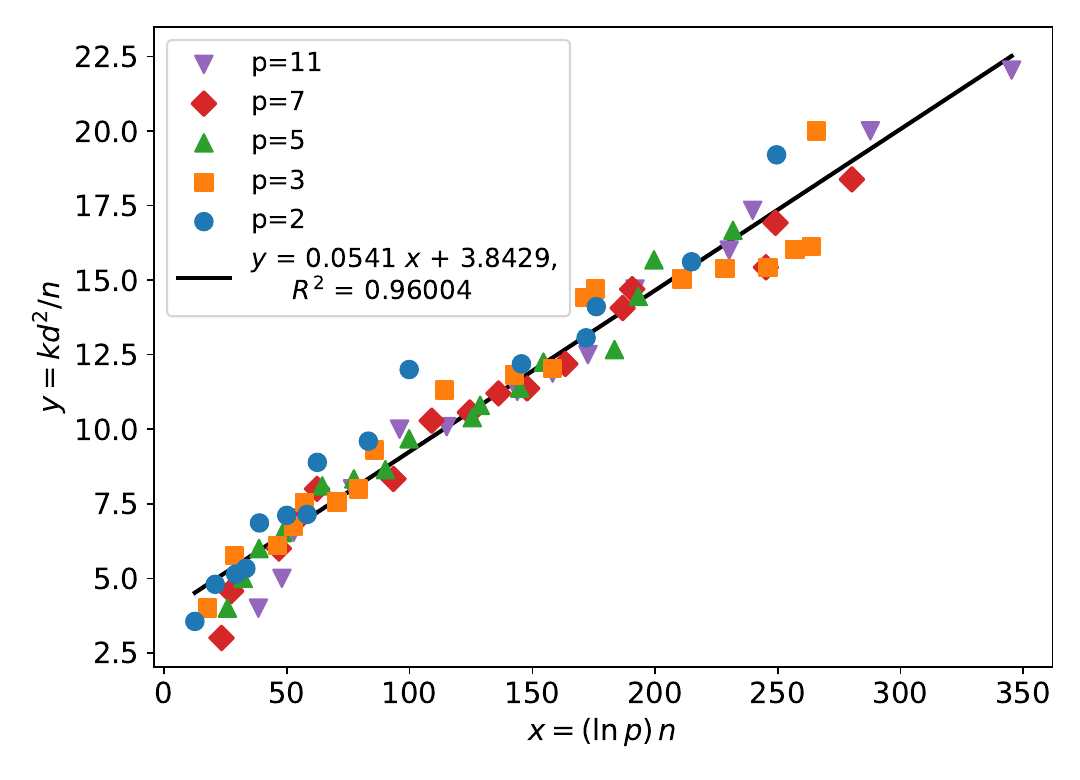}
\caption{Empirical scaling of finite-size code performance.
The metric $kd^2/n$ for the code instances listed in Tables~\ref{tab: n_k_d 1}--\ref{tab: n_k_d 5} is plotted against $(\ln p)n$. Blue circles, orange squares, green upward triangles, red diamonds and purple downward triangles denote $p=2,3,5,7,11$, respectively. Here $n$ and $k$ are the numbers of physical and logical qudits, and $d$ denotes the code distance, using reported upper bounds where exact distances are unavailable. The solid black line is the least-squares fit $kd^2/n=0.0541(\ln p)n+3.84$, with coefficient of determination $R^2=0.96$.}
    \label{fig: kd2_n2_lnp_fit}
\end{figure}

\section{Relation to the Bravyi--Poulin--Terhal Tradeoff}
Motivated by the trend in $k d^{2}/n$ across different primes $p$ in Fig.~\ref{fig: table to figure}, we replot the same data in Fig.~\ref{fig: kd2_n2_lnp_fit} as $k d^{2}/n$ versus $(\ln p)\,n$ and perform a linear least-squares fit, obtaining
\begin{equation}
    k d^{2} = 0.0541 \, n^{2}\ln p + 3.84 \, n,
    \label{eq: empirical formula}
\end{equation}
for the explored primes $p\in\{2,3,5,7,11\}$.
At first glance, the quadratic dependence on $n$ may seem to contradict the Bravyi--Poulin--Terhal (BPT) bound~\cite{bravyi2009no, Bravyi2010Tradeoffs}, which states that for geometrically local stabilizer codes in two dimensions
\begin{equation}
    k d^{2} = O(n),
    \label{eq: original BPT bound}
\end{equation}
where the implicit constant depends on the interaction range $r$ and the on-site Hilbert-space dimension.

To make this dependence explicit, we follow the BPT argument. The lattice is partitioned into an array of $R\times R$ blocks such that each block is correctable, and the block corners are covered by a region $C$. By the cleaning lemma, one may choose $R=\Omega(d/r)$. The proof in Ref.~\cite{Bravyi2010Tradeoffs} then implies
\begin{equation}
    k \le S(C) \le |C|
        = O\!\left(\frac{n r^{2}}{R^{2}}\right)
        = O\!\left(\frac{n r^{4}}{d^{2}}\right),
\end{equation}
where $S(C)$ denotes the entanglement entropy of region $C$. Here, we use the fact that each corner region has linear size $O(r)$ (and hence area $O(r^{2})$), and that the number of blocks is $n/R^{2}$.
It follows that
\begin{equation}
    k d^{2} = O(n r^{4}).
\end{equation}
Superlinear behavior in $n$ is therefore compatible with BPT whenever the interaction range $r$ increases with system size. In our generalized toric constructions, the geometric diameter of the weight-$6$ stabilizers can grow with $n$, depending on the chosen exponents in $f(x,y)$ and $g(x,y)$. This relaxes the BPT constraint and is consistent with the empirical relation~\eqref{eq: empirical formula}.

\section{Discussion and future directions}
We have developed a topological-order perspective on translation-invariant $\ZZ_p$ stabilizer codes on twisted tori.
In this framework, the logical dimension $k$ is determined
by the anyon algebra and computed as twice the
$\mathbb{Z}_p$-vector-space dimension of the quotient ring
$\mathcal{R}/I$, where
$I=\langle f(x,y),g(x,y),y^\alpha-1,x^\beta y^\gamma-1\rangle$
is the ideal incorporating the stabilizer polynomials and
the twisted boundary conditions.
This ring-theoretic approach naturally incorporates twisted boundary conditions and, together with Gr\"obner-basis techniques, enables efficient characterization and systematic searches for new qudit LDPC codes. The high-performing instances summarized in Table~\ref{tab: n_k_d 1} provide a foundation for further theoretical and hardware-oriented investigations.

Future work may extend the search to larger system sizes, where further improvements could emerge. Because the enumeration procedure is fully parallelizable, it can be scaled to $n\le 500$ or beyond using clusters or supercomputers, approaching the scale of current experimental platforms~\cite{Google2019supremacy, Pan2020Quantum, Lukin2021programmable, Madsen2022Quantum, Bravyi2022future, Lukin2025Quantum}. 
{\change
The primary computational bottleneck remains distance estimation.
For all listed instances with reported upper bounds greater
than 15, we use $10^8$ information sets per code.
The convergence tests in
Appendix~\ref{app:distance-convergence} provide evidence for
the reproducibility and stability of the bounds for the twelve
tested instances.
Establishing exact distances and assessing convergence for
larger codes remain important challenges, requiring additional
computational effort and stronger distance-verification methods.
}
It will also be important to verify whether the empirical relation~\eqref{eq: empirical formula} persists at larger $n$ and $p$, and to establish a theoretical understanding of the observed $n^{2}\ln p$ scaling.

{\change
Our framework also allows more general normalized polynomial pairs
\begin{equation}
\begin{aligned}
f(x,y)&=1+r_1x^{a_1}y^{b_1}+r_2x^{a_2}y^{b_2},\\
g(x,y)&=1+r_3x^{c_1}y^{d_1}+r_4x^{c_2}y^{d_2},
\end{aligned}
\end{equation}
with nonzero coefficients and integer exponents. The ansatz in Eq.~\eqref{eq: a b c d generalized toric code} fixes $(a_1,b_1)=(1,0)$ and $(c_1,d_1)=(0,1)$. In our previous study of qubit codes~\cite{liang2025Generalized}, tests of the broader three-term family for $n\leq108$ did not identify improved $[[n,k,d]]_2$ parameters on twisted tori. This observation provides numerical motivation for the restricted ansatz, but does not establish its optimality at larger sizes or for $p>2$. 
More general supports and other term partitions, such as $(\operatorname{wt}(f),\operatorname{wt}(g))=(2,4)$, may yield additional codes and improved parameters. Their number and attainable parameters are not determined by the present search, and our numerical conclusions remain restricted to the explored family and parameter ranges.
}

Furthermore, allowing higher-weight stabilizers is another natural direction, as such constructions have led to improved $[[n,k,d]]$ parameters in other quantum LDPC code families~\cite{Kovalev2013QuantumKronecker, breuckmann2021balanced, panteleev2021degenerate, Lin2022c3Locally, Leverrier2022Tanner, Panteleev2022goodqldpc, Dinur2023Good, wang2023abelian, Lin2024Quantumtwoblock, tiew2024low, Wills2024Localtestability, eberhardt2024logical, Wills2025Tradeoff, liang2026selfdual}. 

Finally, it will be important to assess the performance of these codes under circuit-level noise models~\cite{Bravyi2024HighThreshold}. In particular, syndrome-extraction circuits must be optimized for low depth, as the circuit-level distance is typically smaller than the code distance. Numerical simulations to estimate pseudo-thresholds, along with the development of efficient logical-gate implementations and hardware realizations, will be essential steps toward practical deployment.

\section{Data availability}
The code parameters and summarized numerical results are provided
in this article.
The complete raw datasets generated by the exhaustive searches
within the code family and parameter ranges specified in the main text
are not publicly available owing to their large volume, but are
available from the corresponding author upon reasonable request.

\section{Code availability}
The methods used for probabilistic distance estimation are
described in Ref.~\cite{Pryadko2022GAP}, and those used for
exact distance verification are described in
Ref.~\cite{Spencer_2026}.
The custom code used for code construction, exhaustive searches,
and numerical analysis is available from the corresponding author
upon reasonable request.
The search ranges, computational budgets, and numerical procedures
are described in the main text.

\section{Funding}
Y.-A.C. acknowledges support from the National Natural Science
Foundation of China (Grant No.~12474491) and the Fundamental
Research Funds for the Central Universities, Peking University.
This funding supported the computational resources used for
the numerical searches and analyses.

\section{Author contributions}
Z.L. developed the code construction, carried out the analytical derivations and numerical searches, analyzed the data, and prepared the figures and tables. Y.-A.C. conceived and supervised the project and contributed to the theoretical development and interpretation of the results. Both authors wrote the manuscript, discussed the results, reviewed the manuscript, and approved the final version.

\section{Competing interests}
Zijian Liang is a paid intern at Hangzhou MatriQ Computing Co., Ltd.
The other author declares no competing financial or non-financial
interests.

\bibliography{bib.bib}

\bigskip
\onecolumngrid
\appendix
\pagebreak 

\section{$\ZZ_2$ bivariate bicycle codes}\label{app: Z2 BB codes}
This appendix collects representative instances of $\ZZ_2$ bivariate bicycle (BB) codes, compiled from Ref.~\cite{liang2025Generalized}. We adopt the polynomial description used there: each code instance is specified by a pair of Laurent polynomials
$f(x,y)$ and $g(x,y)$ with $\ZZ_2$ coefficients, together with a choice of boundary identifications that render the translation lattice finite. The latter are encoded by two twisted boundary vectors $\vec a_1,\vec a_2\in\ZZ^2$.

For each choice of $(f,g;\vec a_1,\vec a_2)$, one obtains a qubit LDPC stabilizer code with parameters $[[n,k,d]]_2$, where $k$ is the number of logical qubits and $d$ is the code distance of the underlying stabilizer code. Since our goal here is to provide concrete test cases spanning a range of performance, we select examples with progressively larger values of the figure of merit $k d^2/n$. This quantity is a convenient summary statistic that combines rate and distance for finite-size comparisons; it does not by itself determine performance under a specific decoder or circuit-level noise model, but it is useful for organizing representative instances.

Table~\ref{tab: n_k_d 5} lists the resulting code parameters, the defining polynomials $f(x,y)$ and $g(x,y)$, the boundary vectors $\vec a_1,\vec a_2$, and the corresponding $k d^2/n$ values.
These instances serve as representative benchmarks in the main text for comparing finite-size behavior across different code constructions.

\begin{table}[h]
\centering
{\renewcommand{\arraystretch}{1.5}%
\begin{tabular}{|c|c|c|c|c|c|}
\hline
$[[n,k,d]]$ & $~f(x,y)~$ 
& $~g(x,y)~$ & $\vec{a}_1$   &$\vec{a}_2$   
&$\frac{kd^2}{n}$ 
\\ \hline

$[[18,4,4]]$ & $1+x+xy$&$1+y+xy$ &
$(0,3)$&$(3,0)$   
&{3.56} 
\\ \hline

$[[30,4,6]]$ & ${1+x+x^2}$&${1+y+x^2}$ &
$(0,3)$&$(5,1)$   
&{4.8} 
\\ \hline

$[[42,6,6]]$ & ${1+x+xy}$&${1+y+xy^{-1}}$ &
$(0,7)$&$(3,2)$   
&{5.14} 
\\ \hline

$[[48,4,8]]$ & ${1+x+x^2}$&${1+y+x^2}$ &
$(0,3)$&$(8,1)$   
&{5.33} 
\\ \hline

$[[56,6,8]]$ & ${1+x+y^{-2}}$&${1+y+x^{-2}}$ &
$(0,7)$&$(4,3)$   
&{6.86}
\\ \hline

$[[72,8,8]]$ & ${1+x+x^{-1}y^3}$&${1+y+x^3y^{-1}}$ &
$(0,12)$&$(3,3)$   
&{7.11} 
\\ \hline

$[[84,6,10]]$ & ${1+x+x^{-2}}$&${1+y+x^{-2}y^2}$ &
~$(0,14)$~ &~$(3,-6)$~   
&{7.14} 
\\ \hline

$[[90,8,10]]$ & ~${1+x+x^{-1}y^{-3}}$~ &${1+y+x^3y^{-1}}$ &
$(0,15)$&$(3,-6)$   
&{8.89} 
\\ \hline

$[[{120,8,12}]]$ & $1+x+x^{-2}y$&$1+y+xy^2$ &
$(0,10)$&$(6,4)$   
&{9.6} 
\\ \hline

$[[144,12,12]]$ & $1+x+{x^{-1}y^{-3}}$&${1+y+x^3y^{-1}}$ &
$(0,12)$&$(6,0)$   
&{12} 
\\ \hline

~$[[{210,10,16}]]$~ & $1+x+x^{-3}y^2$& ~$1+y+x^{-3}y^{-1}$~ &
$(0,21)$&$(5,10)$   
&{12.19} 
\\ \hline

$[[{248,10,18}]]$ & $1+x+x^{-2}y$&$1+y+x^{-3}y^{-2}$ &
$(0,62)$&$(2,25)$   
&~{13.06}~ 
\\ \hline

$[[{254,14,16}]]$ & $1+x+x^{-1}y^{-3}$&$1+y+y^{-6}$ &
$~(0,127)~$&$(1,25)$   
&{14.11} 
\\ \hline

$[[{310,10,{\change \leq}22}]]$ & $1+x+x^3y^2$&$1+y+x^{-4}y^4$ &
$(0,31)$&$(5,11)$   
&{{\change $\leq$}15.61} 
\\ \hline

$~[[360,12,{\change \leq}24]]$~  &$1+x+{x^{-1}y^3}$&${1+y+x^3y^{-1}}$&
$(0,30)$&$(6,6)$   
&{{\change $\leq$}19.2}
\\ \hline

\end{tabular}}
\caption{$\ZZ_2$ qubit LDPC codes compiled from Ref.~\cite{liang2025Generalized}. The table reports the code parameters $[[n,k,d]]$, the corresponding polynomials $f(x,y)$ and $g(x,y)$, the twisted boundary vectors $\vec{a}_1,\vec{a}_2$, and the metric $k d^{2}/n$.}
\label{tab: n_k_d 5}
\end{table}

\section{Convergence of the probabilistic distance search}
\label{app:distance-convergence}

\begin{figure}[t]
    \centering
\includegraphics[width=0.8\linewidth,trim=0 0 0 22bp,clip]{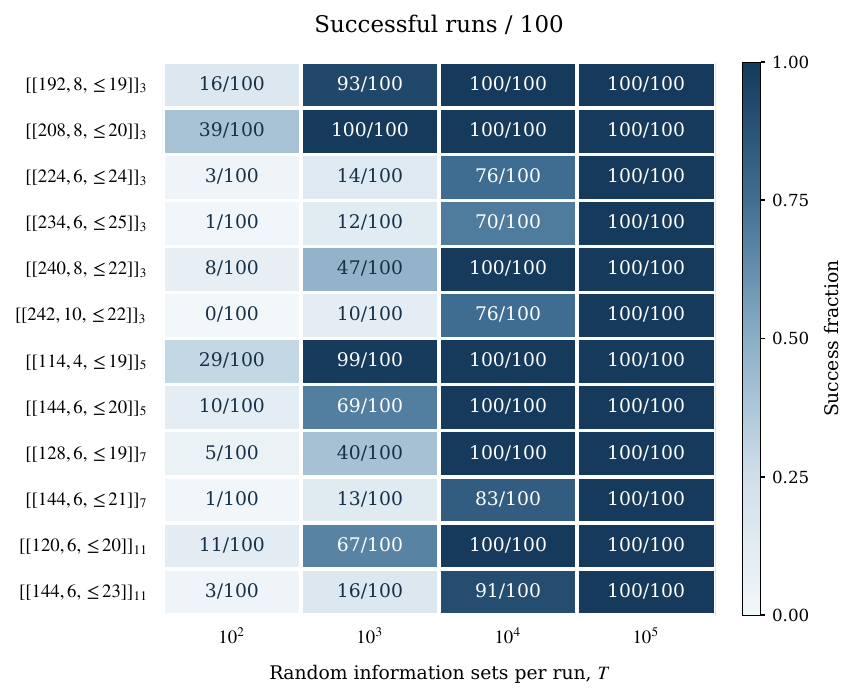}
    \caption{
    {\change
    Convergence tests of the randomized information-set search
    for the twelve qudit codes with reported distance upper bounds
    of at least 19.
    For each code, 100 independently seeded runs are evaluated after
    $T=10^2,10^3,10^4,10^5$ information sets per run.
    A run is successful if the smallest logical-operator weight
    found does not exceed the upper bound indicated in the row label.
    Each cell displays the number of successful runs out of 100;
    the color encodes the corresponding fraction.
    The same runs are continued through all checkpoints, and
    all 100 runs succeed for every code by $T=10^5$.
    Separately, every entry in Table~\ref{tab: n_k_d 1} with a reported upper bound
    greater than 15 is searched using a cumulative total of
    $10^8$ information sets per code.
    For the twelve codes shown here, these extended searches
    return the bounds displayed in the row labels, with no
    smaller value found.
    }}
    \label{fig:distance-convergence}
\end{figure}

{\change
To assess the reliability of the reported distance upper bounds,
we performed convergence tests on the twelve qudit code instances
with reported bounds of at least 19, covering $p=3,5,7,11$.
We examined how reproducibly independent randomized searches
attain these bounds and whether a larger search budget yields
smaller logical-operator weights.

For each of these twelve codes, we performed 100 independently
seeded runs of the randomized information-set search, recording
the smallest nontrivial logical-operator weight found after
$T=10^2,10^3,10^4,10^5$ information sets per run.
A run was counted as successful at a checkpoint if it met or
improved upon the reported upper bound shown in the row label.
No logical operator found in a previous search was supplied
as an initial witness, and each run continued to $T=10^5$
even if it had already succeeded.
The same 100 runs were continued through all checkpoints;
measurements at different checkpoints are therefore not independent.
The total budget for this repeated-run experiment was
$100\times10^5=10^7$ information sets per code.

Figure~\ref{fig:distance-convergence} shows the number of successful
runs at each checkpoint.
At $T=10^2$, the success fractions ranged from $0/100$ to $39/100$;
at $T=10^3$, they ranged from $10/100$ to $100/100$.
At $T=10^4$, seven of the twelve codes reached $100/100$, while
the remaining five reached between $70/100$ and $91/100$.
By $T=10^5$, all 100 runs succeeded for every code.
These results quantify the variation between independent searches
at smaller budgets and show that the reported bounds were
reproducibly attained as the search budget increased.

In addition to this repeated-run experiment, every instance
in Table~\ref{tab: n_k_d 1} with a reported distance upper bound greater than 15
was searched using a cumulative total of $10^8$ information sets
per code, summed over all parallel runs.
The smallest nontrivial logical-operator weight found in these
searches is reported as the distance upper bound.
This budget applies to each code as a whole, whereas the
horizontal axis in Fig.~\ref{fig:distance-convergence} specifies
the budget per independent run.
For the twelve codes shown in the figure, the searches with
$10^8$ information sets returned the same bounds as those
displayed in the row labels; no smaller value was found.

Together, these tests provide numerical evidence for the
reproducibility and stability of the reported upper bounds.
They complement the exact-distance verification of the listed
instances with reported distances at most 15.
However, agreement across repeated searches and the absence of
an improvement within a finite budget do not exclude lower-weight
logical operators.
Distances not verified by exact calculations, and the corresponding
estimates of $kd^2/n$, therefore remain explicitly identified
as upper bounds.
}

\end{document}